\documentstyle[prl,aps,epsf,floats,twocolumn,graphicx,amssymb]{revtex} % COMMENT OUT IF
\newcommand{\be}{\begin{equation}}
\newcommand{\ee}{\end{equation}}
\newcommand{\lr}{\leftrightarrow}
\begin{document}
\draft
%{}`                            
% COMMENT OUT NEXT LINE IF PREPRINT
\twocolumn[\hsize\textwidth\columnwidth\hsize\csname
@twocolumnfalse\endcsname

\title{Patterns of link reciprocity in directed networks}
\author{Diego Garlaschelli$^{1,2}$, Maria I. Loffredo$^{2,3}$}
\address{$^1$Dipartimento di Fisica, Universit\`a di Siena, Via Roma 56, 53100 Siena ITALY\\
$^2$INFM UdR Siena, Via Roma 56, 53100 Siena ITALY\\
$^3$Dipartimento di Scienze Matematiche ed Informatiche, Universit\`a di Siena, Pian dei Mantellini 44, 53100 Siena ITALY}
\date{\today}
\maketitle
\begin{abstract}
We address the problem of link reciprocity, the non-random presence of two mutual links between pairs of vertices. We propose a new measure of reciprocity that allows the ordering of networks according to their actual degree of correlation between mutual links. We find that real networks are always either correlated or anticorrelated, and that networks of the same type (economic, social, cellular, financial, ecological, etc.) display similar values of the reciprocity. The observed patterns are not reproduced by current models. 
This leads us to introduce a more general framework where mutual links occur with  a conditional connection probability. 
In some of the studied networks we discuss the form of the conditional connection probability and the size dependence of the reciprocity.
\end{abstract}

\pacs{89.75.-k, 05.65.+b}
]
\narrowtext
The recent discovery of a complex network structure in many different physical, biological and socioeconomic systems has triggered an increasing effort in understanding the basic mechanisms determining the observed topological organization of networks\cite{barabba,siam}. Nontrivial properties such as a \emph{scale-free} character, clustering, and correlations between vertex degrees are now widely documented in real networks, motivating an intense theoretical activity concerned with network modelling\cite{barabba,siam,assort}.

In the present paper we focus on a peculiar type of correlation present in directed networks: \emph{link reciprocity}, or the tendency of vertex pairs to form mutual connections between each other\cite{recipr}. In other words, we are interested in determining whether double links (with opposite directions) occur between vertex pairs more or less often than expected by chance. 
This problem is fundamental for several reasons. Firstly, if the network supports some propagation process (such as the spreading of viruses in e-mail networks\cite{email,kiel} or the iterative exploration of Web pages in the WWW\cite{www}), then the presence of mutual links will clearly speed up the process and increase the possibility of reaching target vertices from an initial one. 
By contrast, if the network mediates the exchange of some good, such as wealth in the World Trade Web\cite{serrano,mywtw,data} or nutrients in food webs\cite{havens,nature}, then any two mutual links will tend to balance the flow determined by the presence of each other.
The reciprocity also tells us how much information is lost when a directed network is regarded as undirected (as often done, for instance when measuring the \emph{clustering coefficient}\cite{barabba,email,kiel,serrano,mywtw}).
Finally, detecting nontrivial patterns of reciprocity is interesting by itself, since it can reveal possible mechanisms of social, biological or different nature that systematically act as organizing principles shaping the observed network topology.

In general, directed networks range between the two extremes of a \emph{purely bidirectional} one (such as the Internet, where information always travels both ways along computer cables) and of a \emph{purely unidirectional} one (such as citation networks\cite{barabba}, where recent papers can cite less recent ones while the opposite cannot occur). A traditional way of quantifying where a real network lies between such extremes is measuring its reciprocity $r$ as the ratio of the number of links pointing in both directions $L^\lr$ to the total number of links $L$\cite{recipr,email,serrano}:
\be
r\equiv\frac{L^\lr}{L}
\label{r}
\ee
Clearly, $r=1$ for a purely bidirectional network while $r=0$ for a purely unidirectional one. In general, the value of $r$ represents the average probability that a link is reciprocated.
Social networks\cite{recipr}, email networks\cite{email}, the WWW\cite{email} and the World Trade Web\cite{serrano} were recently found to display an intermediate value of $r$.

However, the above definition of reciprocity poses various conceptual problems that we would like to highlight before proceeding with a systematic analysis of real networks. 
Firstly, the measured value of $r$ must be compared with the value $r^{rand}$ expected in a random graph with the same number of vertices and links in order to assess if mutual links occur more or less (or just as) often than expected by chance\cite{email}. This means that $r$ has only a \emph{relative} meaning and does not carry complete information by itself. 
Secondly, and consequently, the definition (\ref{r}) does not allow a clear ordering of different networks with respect to their actual degree of reciprocity. To see this, note that $r^{rand}$ is larger in a network with larger link density (since mutual connections occur by chance more often in a network with more links), and as a consequence it is impossible to compare the values of $r$ for networks with different density, since they have distinct reference values. 
Finally note that, even in two networks with the same density, the definition (\ref{r}) can give inconsistent results if $L$ includes the number of self-loops (links starting and ending at the same vertex). Since the latter can never occur in mutual pairs, while their number can vary significantly across different networks, a finer measure of reciprocity should exclude them from the potential set of mutual connections (hence $L$ should be defined as the number of links minus that of self-loops).

In order to avoid the aforementioned problems, we propose a new definition of reciprocity (denoted as $\rho$ to avoid confusion with $r$) as the correlation coefficient between the entries of the adjacency matrix of a directed graph ($a_{ij}=1$ if a link from $i$ to $j$ is there, and $a_{ij}=0$ if not):
\be\label{rho1}
\rho\equiv\frac{\sum_{i\ne j}(a_{ij}-\bar{a})(a_{ji}-\bar{a})}{\sum_{i\ne j}(a_{ij}-\bar{a})^2}
\ee
where the average value $\bar{a}\equiv\sum_{i\ne j}a_{ij}/N(N-1)=L/N(N-1)$ measures the ratio of observed to possible directed links (link density), and self-loops are now and in the following excluded from $L$, since $i\ne j$ in the sums appearing in eq.(\ref{rho1}).
Note that with such a choice $r^{rand}=\bar{a}$, since in an uncorrelated network the average probability of finding a reciprocal link between two connected vertices is simply equal to the average probability of finding a link between any two vertices, which is given by $\bar{a}$.

Although the above definition appears much more complicated than eq.(\ref{r}), it reduces to a very simple expression. Indeed, since $\sum_{i\ne j}a_{ij}a_{ji}=L^\lr$ and $\sum_{i\ne j}a^2_{ij}=\sum_{i\ne j}a_{ij}=L$, eq.(\ref{rho1}) simply gives
\be\label{rho2}
\rho=\frac{L^{\leftrightarrow}/L-\bar{a}}{1-\bar{a}}=\frac{r-\bar{a}}{1-\bar{a}}
\ee
The correlation coefficient $\rho$ is free from the conceptual problems mentioned above, since it is an \emph{absolute} quantity which directly allows to distinguish between \emph{reciprocal} ($\rho>0$) and \emph{antireciprocal} ($\rho<0$) networks, with mutual links occuring more and less often than random respectively. 
In this respect, $\rho$ is similar to the \emph{assortativity coefficient} \cite{assort} which allows to distinguish between assortative or disassortative networks. The neutral or \emph{areciprocal} case corresponds to $\rho=0$. Note that if all links occur in reciprocal pairs one has $\rho=1$ as expected. However, if $L^\leftrightarrow=0$ one has $\rho=\rho_{min}$ where
\be\label{rhomin}
\rho_{min}\equiv-\frac{\bar{a}}{1-\bar{a}}
\ee
which is always from $\rho=-1$ unless $\bar{a}=1/2$. This occurs because in order to have perfect anticorrelation ($a_{ij}=1$ whenever $a_{ji}=0$) there must be the same number of zero and nonzero elements of $a_{ij}$, or in other words half the maximum possible number of links in the network. This is another remarkable advantage of using $\rho$, since it incorporates the idea that complete antireciprocity ($L^\lr=0$) is more statistically significant in networks with larger density, while it has to be regarded as a less pronounced effect in sparser networks. Also note that the expression for $\rho_{min}$ only makes sense if $\bar{a}\le 1/2$, since with higher link density it is impossible to have $L^\lr=0$ and the minimum reciprocity is no longer given by eq.(\ref{rhomin}) (values of $\bar{a}$ larger than $1/2$ are observed for the most recent data of the World Trade Web shown below). 
Finally note that the definition (\ref{rho1}) allows a direct generalization to weighted networks or graphs with multiple edges by substituting $a_{ij}$ with any matrix $w_{ij}$.

As in ref.\cite{assort}, we can evaluate the standard deviation $\sigma_\rho$ for $\rho$ in terms of the values $\rho_{ij}$ obtained when any (single or not) link between vertices $i$ and $j$ is removed:
\begin{eqnarray}
\sigma^2_\rho&=&\sum_{i<j}(\rho-\rho_{ij})^2\\
&=&L^\lr(\rho-\rho^\lr)^2+(L-L^\lr)(\rho-\rho^\to)^2\nonumber
\end{eqnarray}
where ${ \rho^\lr=\frac{(L^\lr-2)/(L-2)-(L-2)/N(N-1)}{1-(L-2)/N(N-1)} }$ is the value of $\rho$ when a pair of mutual links is removed and ${ \rho^\to=\frac{L^\lr/(L-1)-(L-1)/N(N-1)}{1-(L-1)/N(N-1)} }$ is the value of $\rho$ when the link between two singly connected vertices is removed.

We can now proceed with the analysis of the reciprocity in a coherent fashion.
Table \ref{tab} shows the values of $\rho$ computed on 133 real networks. The most striking result is that, when ordered by decreasing values of $\rho$ as shown in the table, all networks result clearly arranged in groups of the same kind. The most correlated system is the international import/export network or World Trade Web (WTW), displaying $0.68\le\rho\le 0.95$ for each of its 53 annual snapshots \cite{data} in the time interval 1948-2000 (more details on this system are given below).
The WTW is followed by a portion of the WWW\cite{www} and by two versions of the neural network of the nematode \emph{C. elegans}\cite{neural1,neural2} (one where the vertices are different neuron classes and one where they are single neurons).
For the two neural networks, we find that the reciprocity is preserved ($\rho_{neurons}=0.17\pm 0.02$ and $\rho_{classes}=0.18\pm 0.04$) even after removing the links corresponding to gap junctions (which, differently from the chemical synapses, are intrinsically bidirectional\cite{neural1,neural2}).
We then have two different e-mail networks (one built from the address books of users\cite{email} and one from the actual exchange of messages\cite{kiel} in two different Universities). The little difference in their values of $\rho$ suggests the presence of a similar underlying social structure between pairs of users, either appearing in each other's address book or mutually exchanging actual messages. 
A similar consideration applies to the two word association networks \cite{words} (one based on the relations between the terms of the \emph{Online Dictionary of Library and Information Science} and one on the empirical free associations between words collected in the \emph{Edinburgh Associative Thesaurus}), since completely free associations between words seem to reproduce most of the mutuality present in a network with logically or semantically linked terms, an interesting effect probably related to some intrinsic psychological factor. 
The weakly correlated range $0.006\le\rho\le 0.052$ is covered by the 43 cellular networks of ref.\cite{metab}, where reciprocity is related to the potential reversibility of biochemical reactions.
Finally, we find that the antireciprocal region $\rho<0$ hosts the shareholding networks corresponding to two US financial markets\cite{shares} and 28 different food webs: the 22 largest ones of ref.\cite{havens} and the six ones studied in ref.\cite{nature}.
We note that often $\rho=\rho_{min}$ for both classes of networks, highlighting the tendency of companies to avoid mutual financial ownerships and the scarce presence of mutualistic interactions (symbiosis) in ecological webs.

%%%%%%%%%%%%%%%%%%%%%%%%% Table %%%%%%%%%%%%%%%%%%%%%%%
\begin{table}[ht]
\begin{centering}
\begin{tabular}{|l|c|c|c|}
Network& $\rho$ & $\sigma_\rho$ & $\rho_{min}$ \\
\hline
\hline
\textbf{Perfectly reciprocal} & \textbf{1} & $-$ & ${\bf -\frac{\displaystyle \bar{a}} {\displaystyle 1-\bar{a}} }$\\
\hline
\textbf{World Trade Web} (53 webs)\cite{data}&&&\\
\quad most correlated (year 2000) & 0.952 & 0.002 & $(\bar{a}>.5)$ \\
\quad least correlated (year 1948)& 0.68 & 0.01 & -0.80\\
\hline
\textbf{World Wide Web}\cite{www}& 0.5165 & 0.0006 & -0.0001\\
\hline
\textbf{Neural Networks}\cite{neural1,neural2}&&&\\
\quad Neuron classes & 0.44 & 0.03 & -0.04\\
\quad Neurons & 0.41 & 0.02 & -0.03\\
\hline
\textbf{Email Networks}\cite{email,kiel}&&&\\
\quad Address books & 0.231 & 0.003 & -0.001\\
\quad Actual messages & 0.194 & 0.002 & -0.001\\
\hline
\textbf{Word Networks}\cite{words} &&&\\
\quad Dictionary terms & 0.194 & 0.005 & -0.002\\
\quad Free associations & 0.123 & 0.001 & -0.001\\
\hline
\textbf{Cellular Networks} (43 webs)\cite{metab} & & &\\
\quad most correlated (\emph{H. influenzae}) & 0.052 & 0.006 & -0.001\\
\quad least correlated (\emph{A. thaliana}) & 0.006 & 0.004 & -0.003\\
\hline
\textbf{Areciprocal}& \textbf{0}& $-$& ${\bf -\frac{\displaystyle \bar{a}} {\displaystyle 1-\bar{a}} }$\\
\hline
\textbf{Shareholding Networks} \cite{shares}&&&\\
\quad NYSE & -0.0012 & 0.0001 & -0.0012\\
\quad NASDAQ & -0.0034 & 0.0002 & -0.0034\\
\hline
\textbf{Food Webs}\cite{havens,nature}&&&\\
\quad Silwood Park & -0.0159 & 0.0008 & -0.0159\\
\quad Grassland & -0.018 & 0.002 & -0.018\\
\quad Ythan Estuary & -0.031 & 0.005 & -0.034\\
\quad Little Rock Lake & -0.044& 0.007 & -0.080\\
\quad Adirondack lakes (22 webs): & & & \\
\qquad most correlated (B. Hope) & -0.06 & 0.02 & -0.10 \\
\qquad least correlated (L. Rainbow) & -0.102 & 0.007 & -0.102\\
\quad St. Marks Seagrass & -0.105 & 0.008 & -0.105\\
\quad St. Martin Island & -0.13 & 0.01 & -0.13\\
\hline
\textbf{Perfectly antireciprocal}& $\textbf{-1}$ & 
$-$&$\textbf{-1}$\\
\end{tabular}
\caption{Values of $\rho$ (in decreasing order), $\sigma_\rho$ and $\rho_{min}$ for several networks. For three large groups of networks, only the most and the least correlated ones are shown.}
\label{tab}
\end{centering}
\end{table}
%%%%%%%%%%%%%%%%%%%%%%%%%%%%%%%%%%%%%%%%%%%%%%%%%%%%%%%%%%

This clear ordering of network classes according to their reciprocity 
suggests that in each class there is an inherent mechanism yielding systematically similar values of the reciprocity, or in other words that the reciprocity structure is a peculiar aspect of the topology of various directed networks. 
In all cases we find that real networks are either reciprocal or antireciprocal ($\rho_{real}\ne 0$), in striking contrast with current models that generally yield areciprocal networks ($\rho_{model}=0$). To see this, note that $\rho$ aggregates the information about a deeper mechanism existing between each pair of vertices. Let ${p_{ij}\equiv p(i\to j)}$ denote the probability that a link is drawn from vertex $i$ to vertex $j$. In the general case, the probability $p_{i\lr j}$ of having a pair of mutual links between $i$ and $j$ is given by
\be\label{plr}
p_{i\lr j}\equiv p(i\to j\cap i\gets j)=r_{ij}p_{ji}=r_{ji}p_{ij}
\ee
where $r_{ij}$ is the \emph{conditional probability} of having a link from $i$ to $j$ \emph{given that} the mutual link from $j$ to $i$ is there:
\be\label{rij}
r_{ij}\equiv p(i\to j|j\to i)
\ee
Note that $\langle r_{ij}\rangle\equiv\sum_{i\ne j}r_{ij}/N(N-1)=r$, motivating the choice of the symbol.
The expected value of $\rho$ reads
\be\label{rhor}
\rho=\frac{\sum_{i\ne j}p_{ij}r_{ji}-(\sum_{i\ne j}p_{ij})^2/N(N-1)}
{\sum_{i\ne j}p_{ij}-(\sum_{i\ne j}p_{ij})^2/N(N-1)}
\ee
Now, in most models the presence of the mutual link does not affect the connection probability, or in other words $r_{ij}=p_{ij}$ and $p_{i\lr j}=p_{ij}p_{ji}$. This yields $\rho=0$ in eq.(\ref{rhor}), meaning that model networks are areciprocal.
The only way to integrate reciprocity in the models is considering a nontrivial form ($r_{ij}\ne p_{ij}$) of the conditional probability (hence the information required to generate the network is no longer specified by $p_{ij}$ alone). This allows to introduce, beyond $p_{i\lr j}$, the probability ${p_{i\to j}=p_{ij}-r_{ij}p_{ji}}$ of having a \emph{single} link from $i$ to $j$ (and no reciprocal link from $j$ to $i$), and the probability $p_{i\nleftrightarrow j}$ (fixed by the equality ${p_{i\to j}+p_{i\gets j}+p_{i\leftrightarrow j}+p_{i\nleftrightarrow j}=1}$) of having no link between $i$ and $j$. The network can then be generated by drawing, \emph{for each single vertex pair}, a link from $i$ to $j$, a link from $j$ to $i$, two mutual links or no link with probabilities $p_{i\to j}$, $p_{i\gets j}$, $p_{i\leftrightarrow j}$ and $p_{i\nleftrightarrow j}$ respectively. 

%%%%%%%%%%%%%%%%% fig1 %%%%%%%%%%%%%%%%%%% 
\begin{figure}[]	% in second brace, h=here, t=top, b=bottom
\includegraphics[width=.4\textwidth]{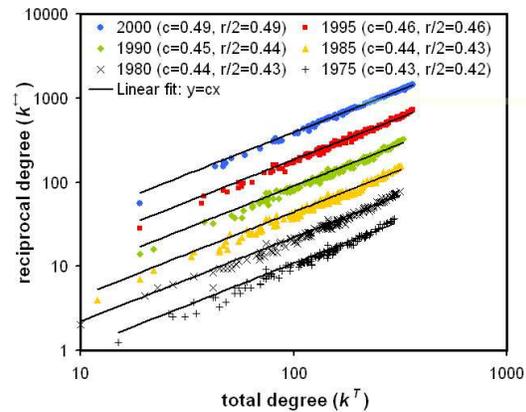}
\caption[]{
\label{fig}
Plots (separated for clarity) of the reciprocal degree $k^\lr$ versus the total degree $k^T$ for six snapshots of the World Trade Web, with linear fit $y=cx$ (error on $c$: $\pm 0.01$).}
\end{figure}
%%%%%%%%%%%%%%%%%%%%%%%%%%%%%%%%%%%%%%%%% 

The form of $r_{ij}$ can be in principle very complicated, however in some of the studied networks we find that it is constant. In particular, we observe that in each snapshot of the World Trade Web the in-degree $k^{in}_i=\sum_j p_{ji}$ and the out-degree $k^{out}_i=\sum_i p_{ij}$ of a vertex are approximately equal, meaning that ${p_{ij}\approx p_{ji}}$ and hence ${r_{ij}\approx r_{ji}}$. Then we find (see fig.\ref{fig}) that for these networks the \emph{reciprocal degree} ${k^\lr_i=\sum_j p_{ij}r_{ji}}$ (number of mutual link pairs of a vertex) is proportional to the total degree ${k^T_i=\sum_j p_{ij}+p_{ji}=2\sum_j p_{ij}}$, or ${k^\lr_i=ck^T_i}$. This means ${r_{ij}\approx r\approx 2c}$, which is confirmed by the excellent agreement between the fitted values of $c$ and the values ${r/2=L^\lr/2L}$ obtained independently (see the legend in fig.\ref{fig}). A similar trend, even if with larger fluctuations, is displayed by the neural networks and the message-based email network (not shown). The other networks instead do not display any clear behaviour, meaning that $r_{ij}$ has in general a more complicated form.

\begin{figure}[]	% in second brace, h=here, t=top, b=bottom
\includegraphics[width=.47\textwidth]{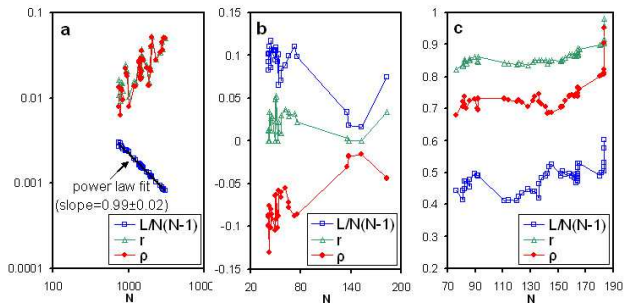}
\caption[]{\label{fig2}
Plots of $\rho(N)$, $r(N)$ and $\bar{a}(N)$ on: \textbf{a)} the 43 cellular networks of ref.\cite{metab}, \textbf{b)} the 28 food webs of refs.\cite{havens,nature} and \textbf{c)} the 53 annual snapshots (1948-2000) of the WTW\cite{data}.}
\end{figure}

Another important problem is the size dependence $\rho(N)$. As evident from eq.(\ref{rho2}), this depends on both $r(N)$ and $\bar{a}(N)$, which display different trends on different classes of networks and therefore should be considered separately for each class. We found three instructive cases, as reported in fig.\ref{fig2}.
For cellular networks ${\bar{a}(N)\propto N^{-1}}$, implying $\rho\to r$ as $N$ increases, therefore the asymptotic behaviour of $\rho$ depends only on that of $r$, which is found to increase as $N$ increases. 
By contrast, $r\approx 0$ for food webs, so that in this case $\rho(N)$ only depends on $\bar{a}(N)$, whose form is however unclear probably due to the small size of the webs \cite{havens}, and therefore no clear trend is observed for $\rho(N)$ as well. The behaviour of the WTW is more complicated because both $r$ and $\bar{a}$ contribute relevantly to $\rho$, and because its $N$-dependence reflects its temporal evolution ($N$ increases monotonically during the considered time interval).
Between 1948 and 1990, $N$ increases from $76$ to $165$ mainly since various colonies become independent states, but $\bar{a}$ and $r$ (and hence $\rho$) fluctuate about a roughly constant value. 
Then, after a sudden increase ($N>180$) in 1991 due to the formation of new states from the USSR, $N$ grows very slowly while $\bar{a}$, $r$ and $\rho$ increase rapidly, an interesting signature of the faster globalization process of the economy and the tighter interdependence of world countries.
Indeed, the steep increase $\rho\to 1$ signals that the world economy is rapidly evolving towards an  ``ordered phase" where all trade relationships are bidirectional.

More generally, this could suggest to promote $\rho$ as an \emph{order parameter} whose continuous variation from $\rho<1$ to $\rho=1$ corresponds to a discontinuous change in the symmetry properties of the adjacency matrix (from a non-symmetric phase to a symmetric, maximally ordered one), a typical behaviour displayed within the theory of second-order phase transitions and critical phenomena. 
The most disordered phase corresponds instead to $\rho=0$, since $r_{ij}=p_{ij}$ and the knowledge of the event $j\to i$ adds no information on the event $i\to j$. The point $\rho=-1$ is again, even if not completely, informative since $r_{ij}=0$.

%%%%%%%%%%%%%%%%% fig2 %%%%%%%%%%%%%%%%%%% 
%%%%%%%%%%%%%%%%%%%%%%%%%%%%%%%%%%%%%%%%% 
The results discussed here represent a first step towards characterizing the reciprocity structure of real networks and understanding its onset in terms of simple mechanisms. Our findings show that reciprocity is a common property of many networks, which is not captured by current models. Our framework provides a preliminary theoretical approach to this poorly studied problem.

We thank Prof. K. Kawamura for kindly providing the data of the \emph{C. elegans} neural network and G. Caldarelli for helpful discussions.

\end{document}